\documentclass[sn-nature]{sn-jnl}

\usepackage{graphicx}%
\usepackage{multirow}%
\usepackage{amsmath,amssymb,amsfonts}%
\usepackage{amsthm}%
\usepackage{mathrsfs}%
\usepackage[title]{appendix}%
\usepackage{xcolor}%
\usepackage{textcomp}%
\usepackage{manyfoot}%
\usepackage{booktabs}%
\usepackage{algorithm}%
\usepackage{algorithmicx}%
\usepackage{algpseudocode}%
\usepackage{listings}%

\theoremstyle{thmstyleone}%
\newtheorem{theorem}{Theorem}
\newtheorem{proposition}[theorem]{Proposition}%

\theoremstyle{thmstyletwo}%
\newtheorem{example}{Example}%
\newtheorem{remark}{Remark}%

\theoremstyle{thmstylethree}%
\newtheorem{definition}{Definition}%

\def\figname{Fig.}
\def\eqname{Eq.}
\def\SiPhName{LightIn }

\begin{document}

\title[Article Title]{Versatile silicon integrated photonic processor: a reconfigurable solution for next-generation AI clusters}



\author[1]{\fnm{Ying} \sur{Zhu}}\email{zhuying@noeic.com}
\author[1]{\fnm{Yifan} \sur{Liu}}\email{liuyifan@noeic.com}

\author[1]{\fnm{Xinyu} \sur{Yang}}\email{yangxinyu@noeic.com}

\author[1,2]{\fnm{Kailai} \sur{Liu}}\email{liukai.lai@outlook.com}

\author[1]{\fnm{Xin} \sur{Hua}}\email{huaxin@noeic.com}

\author[2]{\fnm{Ming} \sur{Luo}}\email{mluo@cict.com}

\author[1]{\fnm{Jia} \sur{Liu}}\email{liujia@noeic.com}

\author[1]{\fnm{Siyao} \sur{Chang}}\email{changsiyao@noeic.com}

\author[1]{\fnm{Jie} \sur{Yan}}\email{yanjie@noeic.com}

\author[1]{\fnm{Shengxiang} \sur{Zhang}}\email{zhangshengxiang@noeic.com}

\author[1]{\fnm{Miao} \sur{Wu}}\email{wumiao@noeic.com}

\author[1]{\fnm{Zhicheng} \sur{Wang}}\email{wangzhicheng@noeic.com}

\author[1]{\fnm{Hongguang} \sur{Zhang}}\email{zhanghongguang@noeic.com}

\author[1]{\fnm{Daigao} \sur{Chen}}\email{chendaigao@noeic.com}

\author*[1,3]{\fnm{Xi} \sur{Xiao}}\email{xiaoxi@noeic.com}

\author[3]{\fnm{Shaohua} \sur{Yu}}\email{yush@cae.cn}

\affil*[1]{\orgdiv{National Information Optoelectronic Innovation Center}, \orgname{China Information and Communication Technologies Group Corporation (CICT)}, \orgaddress{\street{Youkeyuan Road 88}, \city{Wuhan}, \postcode{430074}, \state{Hubei}, \country{China}}}

\affil[2]{\orgdiv{State Key Laboratory of Optical Communication Technologies and Networks}, \orgname{China Information and Communication Technologies Group Corporation (CICT)}, \orgaddress{\street{Gaoxinsi Road 6}, \city{Wuhan}, \postcode{430074}, \state{Hubei}, \country{China}}}

\affil[3]{\orgdiv{Peng Cheng Laboratory}, \orgname{} \orgaddress{\street{Shahexi Road 6001}, \city{Shenzhen}, \postcode{430074}, \state{Guangdong}, \country{China}}}


\abstract{
The Artificial Intelligence models pose serious challenges in intensive computing and high-bandwidth communication for conventional electronic circuit-based computing clusters.
Silicon photonic technologies, owing to their high speed, low latency, large bandwidth, and complementary metal-oxide-semiconductor compatibility, have been widely implemented for data transfer and actively explored as photonic neural networks in AI clusters.
However, current silicon photonic integrated chips lack adaptability for multifuncional use and hardware-software systematic coordination.
Here, we develop a reconfigurable silicon photonic processor with $40$ programmable unit cells integrating over $160$ component, which, to the best of our knowledge, is the first to realize diverse functions with a chip for AI clusters, from computing acceleration and signal processing to network swtiching and secure encryption.
Through a self-developed automated testing, compilation, and tuning framework to the processor without in-network monitoring photodetectors, we implement $4\times4$ dual-direction unitary and $3\times3$ uni-direction non-unitary matrix multiplications, neural networks for image recognition, micro-ring modulator wavelength locking, $4\times4$ photonic channel switching , and silicon photonic physical unclonable functions.
This optoelectronic processing system, incorporating the photonic processor and its software stack, paves the way for both advanced photonic system-on-chip design and the construction of photo-electronic AI clusters.
}




\maketitle

\section{Introduction}\label{sec1}

Artificial intelligence (AI) models are flourishing and demonstrating human-competitive performance in diverse fields, including natural language processing (NLP) \cite{chang2024survey}, computer vision \cite{singh2022flava,wang2024sam},
healthcare \cite{moor2023foundation,huang2023chatgpt}, finance \cite{pallathadka2023applications,cao2024man}, education \cite{chan2023comprehensive}, autonomous driving\cite{zhang2022ai}, scientific research\cite{demszky2023using,chen2024machine,kochkov2024neural}, creative industries\cite{anantrasirichai2022artificial},and more.
These remarkable intelligent capabilities are underpinned by large-scale computational resources processing vast amounts of data, often in petabytes or even exabytes of training data and model parameters \cite{pan2021facebook,zhao2022understanding}.
To meet these computational demands, current AI computing centers have evolved from clusters of thousands of Graphics Processing Units (GPUs) to large-scale systems comprising hundreds of thousands of accelerators \cite{musk2024,oracle2024}.
However, the conventional digital clock-based computing hardware faces challenges due to the slowdown of Moore's Law\cite{mehonic2022brain} and the von-Neurmann architecture bottleneck \cite{filipovich2022silicon,gholami2024ai}.
Consequently, these challenges necessitate the exploration of novel computing architectures and hardware solutions.

Silicon photonics has emerged as a promising solution to these challenges. It offers unique advantages in its complementary metal-oxide-semiconductor (CMOS) compatibility, high speed, low latency, and large bandwidth.
Silicon photonic systems-on-chip have demonstrated superior performance in various applications, including photonic communication\cite{10633778, Shi22}, switching\cite{Seok19}, computing \cite{huang2021silicon,shastri2021photonics,ashtiani2022chip}, and sensing \cite{rogers2021universal}.
Silicon photonic transceivers have become the mainstream solutions in the domain of intra- and inter-datacenter interconnects\cite{shekhar2024roadmapping}.
For shorter distances, optical input/output (I/O) achieves $4$T/b signal transfer with only $5$ns latency and $5$pJ/bit, demonstrating $10\times$ better performance in both speed and energy efficiency compared to electrical I/O \cite{wade2023driving}. 
Most notably for AI applications, recent developments in emerging photonic computing have shown remarkable progress.
The large-scale photonic chiplet Taichi, for instance, achieves $160$TOPS/W energy efficiency for AI acceleration \cite{xu2024large}, demonstrating the significant potential of silicon photonics in advancing AI computing capabilities. A scalable photonic integrated circuit has optically computes both linear and nonlinear functions with a latnecy of $410$ps, which integrate multiple coherent optical processors units for both linear and nonlinear functions into one chip \cite{bandyopadhyay2024single}.

While these application-specific photonic integrated circuits (ASPICs) demonstrate impressive performance, they are inherently limited by their fixed functionality.
The development of ASPICs typically requires multiple design-fab-packaging-test iterations, with each iteration taking several months and incurring substantial costs \cite{perez2017multipurpose,bogaerts2020programmable}.
To overcome these limitations, researchers have proposed photonic field programmable arrays (PFPAs) or general-purpose processors, drawing inspiration from field programmable gate arrays (FPGAs) and central processing units (CPUs) in the electronic domain.
They promise to combine high performance (low cost, compactness, and power efficiency) and rapid and economical functional verification and upgradability \cite{perez2018programmable,bogaerts2020programmable}.
Current implementations include two significant architectures: forward-only and recirculating \cite{bogaerts2020programmable}.
The forward-only architectures primarily employ Mach-Zehnder Interferometer(MZI)-based triangular mesh \cite{reck1994experimental} and rectangular mesh\cite{clements2016optimal,perez2017silicon}.
They enable unitary transformation from multiport inputs to outputs, supporting applications like quantum information processing \cite{harris2017quantum, wang2020integrated}, neurmorphic computing \cite{shen2017deep,harris2018linear}, mode convertion \cite{annoni2017unscrambling}, signal processing \cite{choutagunta2019adapting}. An improved version combining delay lines support multile functions in microwave photonics \cite{xie2023low}. Hoewever they lacks loop routing for infinite impulse response (IIR) filters commonly used in signal processing and control system. 
The recirculating waveguide meshes, with triangular, square, or hexagonal forms, enable both IIR and finite impulse response (FIR) filters \cite{zhuang2015programmable}. 
A hexagonal mesh comprising $72$ programmable unit cells, successfully implements key functions required in 5G/6G wireless systems, such as photonic and RF-photonic filtersing, phase shifting, millimeter-wave generation, tunable delay lines, beamforming, frequency measurement, and optoelectronic oscillators\cite{perez2024general}.
A $9$-cell square mesh dmonstrates fractional differentiation, Hilbert transformation, temporal integrating, routing, and matrix multiplications\cite{zhao2023chip}. 
However, the self-reconfigurable training for scaling and more AI hardware applications remain to be fully explored. 

Here, we demonstrate a silicon-integrated programmable photonic processor based on a $4\times4$ square recirculating mesh with $40$ programmable unit cells (PUCs), representing one of the largest square recirculating mesh implemented to date.
To achieve efficient and stable programming and control, we develop an automatic testing, programming, and calibration (TPC) framework.
By incorporating the programmable photonic processor, an electronic control module, and the TPC framework, we establish a comprehensive prototype processing system, named \SiPhName, to realize diverse functions for AI computing clusters from computing acceleration and signal processing to channel switching and information security.
The processing system has achieved $4\times4$ bidirectional unitary and $3\times3$ non-unitary matrix multiplications based on the universal multiport interferometer structure and the diamond MZI structure, respectively.
We implement it as a neural network to perform an image-recognition task with the automatic TPC framework, achieving competitive accuracy compared to the electronic counterpart.
For signal transmission within the cluster, we configure the processor as differentiators to detect power fluctuations, assisting in wavelength locking for the microring modulators for different baud-rate symbols.
Under photonic differentiator assistance, the microring modulators achieve $32$Gbit/s NRZ modulation with an Extinction Ratio (ER) of $5$dB.
To enable signal switching among multiple computing and storage hardware,$4\times4$ channel switching is programmed on this processor, with crosstalk lower than $20$dB over the $2.5$THz range.
For the information security, we realize physical unclonable functions (PUFs) on the \SiPhName, achieving an intra-die Hamming distance of $1.7\%$ (experimental test) and an inter-die Hamming distance of $50.15\%$ (numerical analysis due to lack of multiple photonic processors).
These realized functions indicate that \SiPhName can provide low-latency and high-power efficiency solutions for high-performance AI computing clusters with a short design and development period and quick function verification.
Furthermore, the proposed automatic TPC framework and potential scalability of the programmable photonic processor lay a solid foundation for the developments and applications of large-scale integration photonic system-on-chips.

\section{Results}
\subsection{Prototype system: architecture and control}
\begin{figure}
\centering
\includegraphics[width=\textwidth]{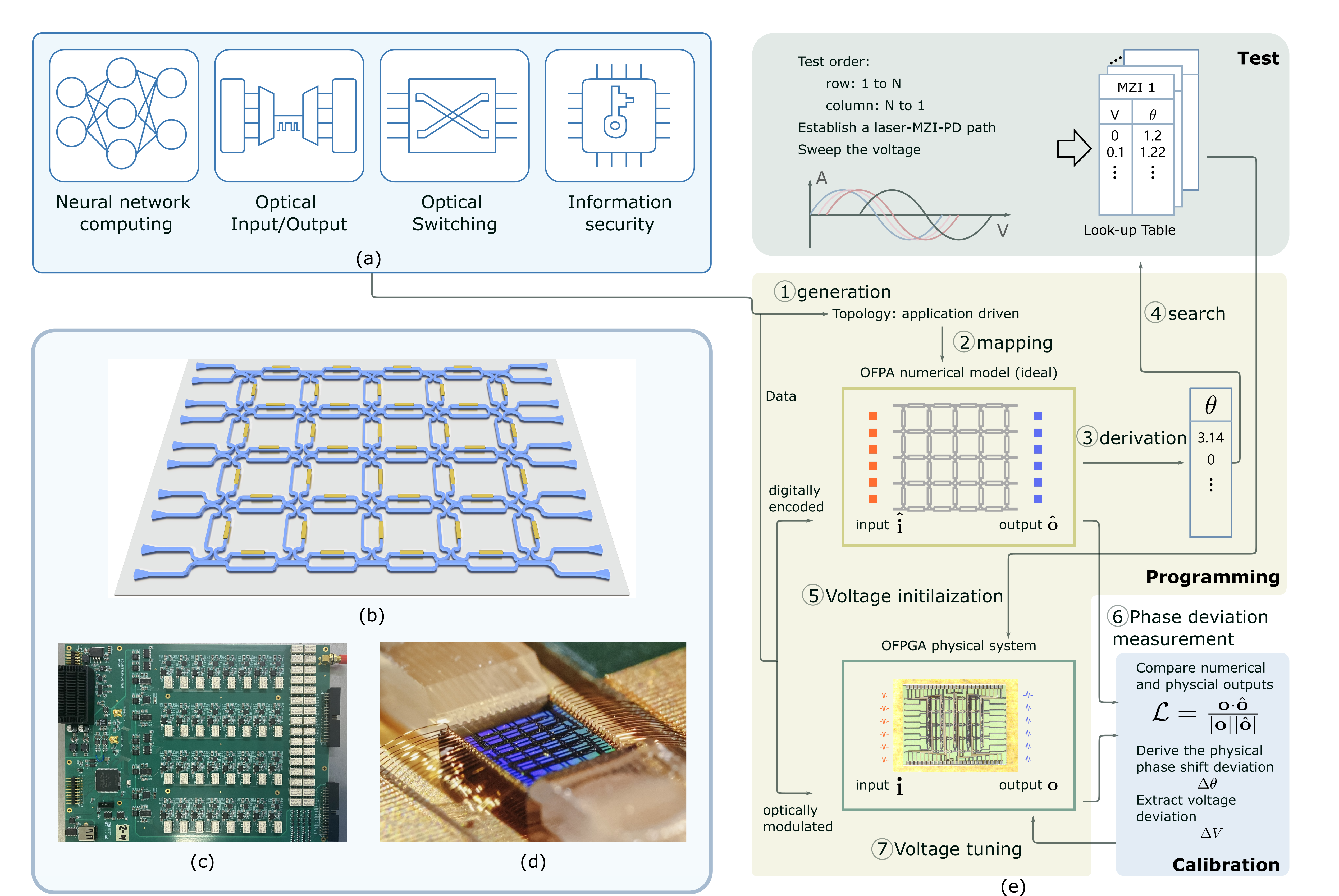}
\caption{The silicon photonic processing system, \SiPhName,and its applications in AI computing clusters. (a) The applications in AI computing clusters. (b) The silicon-integrated programmable photoinc processor conceptual diagram. (c) The electronic control module. (d) The processor packaging and optical coupling setup. (e) The testing, compilation, and adjustment (TCA) framework.}
\label{fig:main}
\end{figure}

The \SiPhName prototype system consists of a silicon photonic processor (see Section \ref{ssec:photonic}) and an electronic control module at the hardware level, complemented by a TPC framework at the software level, as shown in \figname~\ref{fig:main}.
The processor, fabricated using silicon-on-insulator (SOI) technology, features $40$ programmable unit cells (PUCs) arranged in a flat $4\times4$ MZI square mesh topology (\figname~\ref{fig:main}).
$20$ optical ports are equally distributed at the two opposite edges of the processor for signal input/output via two fiber arrays.
Each PUC consists of an MZI with a thermo-optic phase shifter $\theta$ on one arm, whose transformation matrix is
\begin{equation}
\begin{aligned}
T_{PUC} &= \frac{\sqrt{2}}{2} \begin{bmatrix}1&j \\j&1\end{bmatrix}\begin{bmatrix}e^{j\theta}&0 \\0&1\end{bmatrix}\frac{\sqrt{2}}{2} \begin{bmatrix}1&j \\j&1\end{bmatrix}  \\
&=je^{j\frac{\theta}{2}}\begin{bmatrix}\sin\frac{\theta}{2}&\cos\frac{\theta}{2}\\\cos\frac{\theta}{2}&-\sin\frac{\theta}{2}\end{bmatrix}.
\end{aligned}
\label{eq:tm}
\end{equation}
By applying power to $\theta\in$ PUCs, the PUC and the PUC-based square mesh transformation matrices can be configured to process versatile functions.
The configuration is implemented by the electronic control module.
It interfaces with the silicon photonic processor via a Printed Circuit Board (PCB), to which the PUCs are electronically connected through wire bonding.
Details of the prototype processing system and the experimental setup are provided in Section \ref{ssec:prototype}.

To systematically control the \SiPhName processor, we designed and implemented a three-phase test-compile-adjust (TCA) framework (\figname~\ref{fig:main}) within the electronic control module, detailed as follows:

\textbf{Testing: MZI Characteriztion}.
The testing protocol progressively characterizes all MZIs in the mesh through alternating detection and locking.
The testing order is from the first row to the last row and from the last column to the first column as in
\figname~\ref{fig:main}.
For each MZI:
(1) Detection for MZI states:
Sweep the control voltage, measure output intensities, and derive programmed phase shifts via \eqname~\ref{eq:tm}, and build the the voltage-to-phase Look-Up Table (LUT). The cross and bar states are identified from intensity extrema.
(2) Locking MZI states for path generation:
Program pre-tested MZIs to cross/bar states according to their LUTs, establishing a unidirectional optical path from the tested MZI to the output, which can be ensured by the testing order, to provide intensity measurement directly correlate with the tested MZI's phase-voltage response.
(While the untested MZI states are indefinite due to the manufacturing variations, paths generally exist from the input to the tested MZI.)
Repeat the Detection step until characterizing all MZIs.
This hierarchical approach enables stable initialization for subsequent phases.

\textbf{Compilation: Programming voltage initilization}.
The compilation phase determines the MZI phase values and the corresponding initialization voltages required by the tasks.
(1) Topology selection: deploy the predetermined MZI mesh according to the tasks, determining the routing MZIs and functional MZIs.
For example, unitary matrix implementations in MZI-based photonic chips conventionally utilize the rectangle mesh\cite{clements2016optimal}, determining the vertical and edge horizontal MZIs and parts of the horizontal MZIs in the square mesh are the routing MZIs and functional MZIs, respectively, as in \figname~\ref{fig:computing}.
(2) Phase shifter calculation: calculate the phase values according to the predetermined MZI mesh.
For the unitary matrices, phase values of the routing MZIs are fixed to $0$ or $\pi$ for the bar or cross states.
The phase values of the functional MZIs are obtained by decomposing the target unitary matrix.
The programming voltages are searched from the LUTs from the testing phase.
Additional topologies and corresponding phase values supporting signal processing, network switching, and secure encryption are demonstrated in the following sections.

\textbf{Adjustment: Adjoint Calibration}.
The adjustment phase aims to mitigate multi-disturbances: the $\pi$-phase ambiguity caused by the intensity detection, thermal crosstalk during programming, and environmental noises.
We establish an adjoint calibration method by constructing a digital-twin square mesh numerical model and comparing the simulated output-to-input responses ($\hat{r}$) and measured ones ($r$) by $\mathcal{L} = r\cdot\hat{r}/(|r|\cdot|\hat{r}|)$.
To resolve the $\pi$-phase ambiguity, we iteratively program the initialization voltages or plus $V_\pi$ (increasing $\pi$ shifts) to the MZIs to obtain minimal $\mathcal{L}$.
Afterward, tune the programming voltages online according to their gradients to $\mathcal{L}$, obtained via voltage adjustments and observations of $\mathcal{L}$ changes.
The adjustment is significant for the phase-sensitive computing acceleration function.

The TCA framework provides a systematic approach for the following experiments and practical applications of the \SiPhName to AI computing clusters.

\subsection{Computing acceleration in AI clusters}
Many works have indicated that the silicon photonic MZI-based triangular mesh \cite{reck1994experimental} and rectangular mesh\cite{clements2016optimal} show promising in the high power efficient and low latency computing acceleration for neural networks\cite{shen2017deep,zhang2021optical, pai2023experimentally}.
However, they belong to ASPIC chips, enabling constrained uni-directional unitary transmission.
Here, we program the \SiPhName to realize bidirectional unitary matrix multiplication, non-unitary matrix multiplication, and neural networks.

\begin{figure}
\centering
\includegraphics[width=\textwidth]{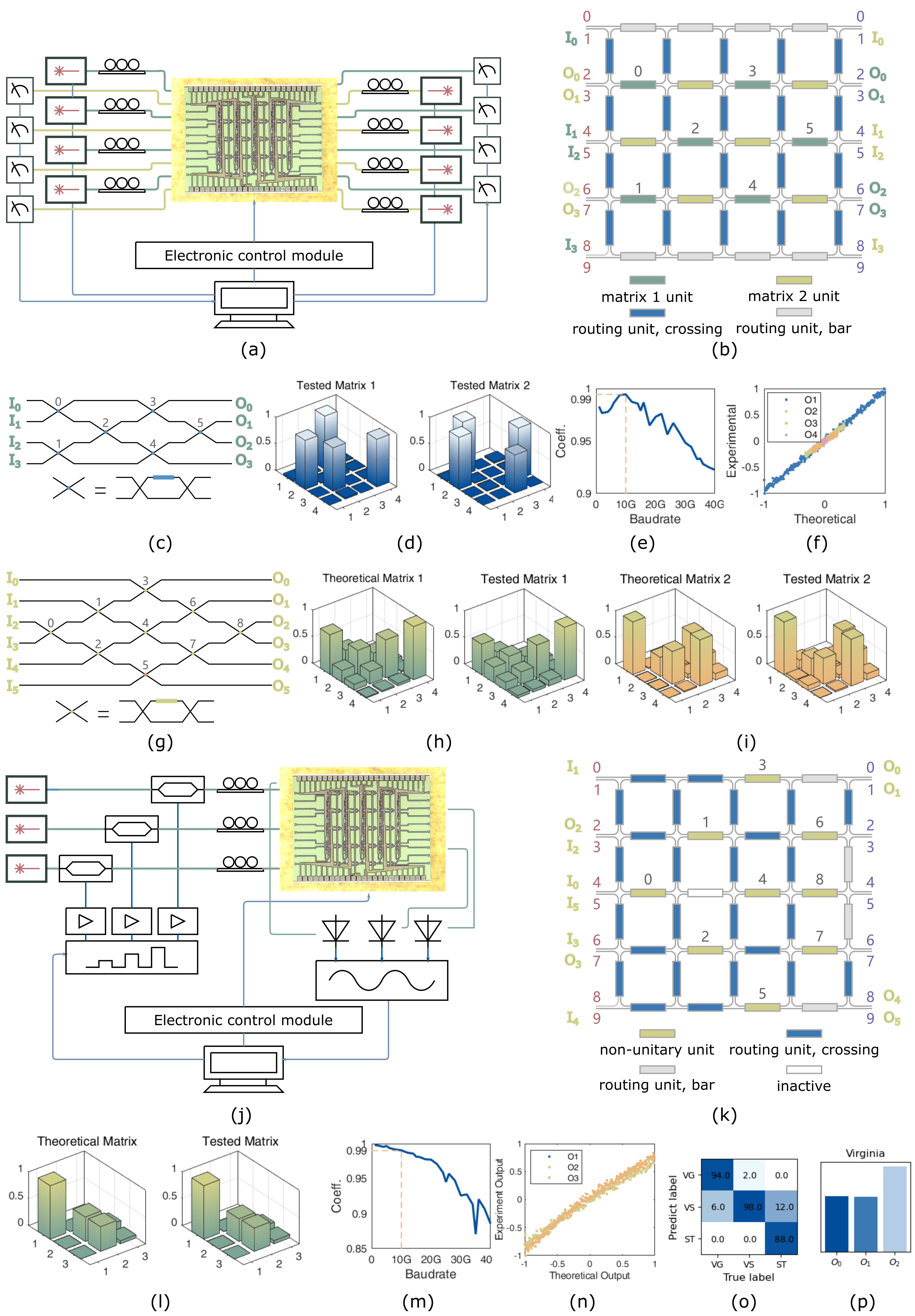}
\caption{Experiment system and results for optical computing implemented on \SiPhName.}
\label{fig:compute}
\end{figure}

\textbf{Bidirectional unitary matrix multiplication}.
\figname~\ref{fig:compute}(a) demonstrates the programming topology of \SiPhName for bidirectional unitary matrix multiplication.
The programmed silicon photonic processor comprises four categories of PUCs: matrix$1$ units (orange), matrix$2$ units (green), forward-only routing units(blue and gray).
This interleaved programming and routing multiplexing improves the footprint efficiency compared to the hexagonal topology by $181.46\%$.

The first experiment is the realization for two $4\times 4$ unitary matrices whose elements are either $1$ or $0$.
With the TPC framework, two matrices are programmed to the photonic processor, one of which is the transmission matrix of the programmable photonic processor from left to right, and the other is from right to left as in \figname~\ref{fig:compute}(a).
We input two $4\times4$ identity matrices from the input ports for both directions, each column of which is sent sequentially.
The corresponding outputs are representative of the two transmission matrices.
\figname~\ref{fig:compute}(b) presents unitary matrix $1$ $\left[\left[0,1,0,0\right],\left[0,0,1,0\right],\left[0,1,0,0\right],\left[0,0,0,1\right]\right]$ and unitary matrix $2$ $\left[\left[0,1,0,0\right],\left[1,0,0,0\right],\left[0,0,0,1\right],\left[0,0,1,0\right]\right]$ programmed on the processor's PUCs, which shows a high fidelity.

Additionally, we implement two random-generated unitary matrices on \SiPhName.
\figname~\ref{fig:compute}(c) illustrates the modulus of the elements in the unitary matrices, which aligns well with their theoretical values.
It is important to note that, due to the lack of coherent detection, intensity detection can only measure the squared modulus of elements for the unitary matrices.
Consequently, when a column vector is input, the output intensities represent the squared modulus multiplication between the vector and unitary matrices.
Nevertheless, we input $256$ trials of different $4\times1$-column vectors to the processor.
The vector values are represented by the amplitudes of non-return-to-zero (NRZ) pulses at a speed of $10$GHz.
The normalized output intensities and theoretical values are depicted in \figname~\ref{fig:compute}(d), presenting a high correlation of $0.99$ ($sigma^2=0.0012$, corresponding to 10.70 bit).
The computing speed achieves $1.92$TOPS, and the averaged energy efficiency is $1.875$pJ/OPS.
\figname~\ref{fig:compute}(e) shows the correlations between the theoretical and experimental computing results under different data baud rates.
We can see that as the baud rate increases, the correlation decreases, meaning the computing resolution decreases.
It could be not caused by the MZI-mesh bandwidth but by the limited AWG sampling rate, which generates the modulation NRZ pules, and the wavelength sensitivity of the grating couplers for input/output of the silicon-integrated programmable photonic processor.
In addition, adding phase shifters to the other arm inter couples of the MZI reduces the power consumption to express negative phase values instead of applying a high power cost to program $2\pi-\theta$, thereby improving energy efficiency \cite{hamerly2024towards}.
Besides, the lack of a phase shift at one input port of the PUC constrains the expression of unitary matrices.
In future work, we will introduce the on-chip modulation for synchronization and coherent detection to obtain the phase information.
We will also update the PUC design by incorporating four phase shifters in both arms at the input and inter-coupler positions.
These allow arbitrary unitary matrix realization and high energy efficiency programming.
More details about the experimental system construction are provided in \ref{ssec:prototype}.

\textbf{Non-unitary matrix multiplication}.
Most matrices for matrix multiplications are non-unitary, which in the previous work is singular-value decomposed into two unitary and one dialog matrices processed by two MZI-based triangle or rectangle meshes and a group of parallel MZIs, amplifiers, or attenuators, respectively.
\figname~\ref{fig:compute}(g) demonstrates a diamond structure capable of expressing general (i.e., non-unitary) matrices, which has advantages of uniform layout and straightforward programming procedure\cite{tischler2018quantum,hamerly2024towards}.
The mathematical derivation for implementing non-unitary operations using the unitary diamond mesh is described in Supplementary Information Note 1.
While due to processor size limitations, a completely forward-only diamond mesh can not be established on the core, more than one available topology exists on the square mesh to realize a $3\times3$ non-unitary matrix with the diamond structure.
We can fold the diamond structure with some column as the axis, and locate the MZIs after the column of the diamond structure to the topologically equivalent MZIs in the square mesh as shown in as shown in \figname~\ref{fig:compute}(g).
This characteristic demonstrates that the processor has not only flexibility but also improves footprint efficiency and fault tolerance.
To validate the design, we implement a randomly generated $3\times3$ non-unitary matrix on \SiPhName and test it with $256$ trials of different $3\times1$-column vectors, encoded by the amplitudes of RZ pulses at $10$GHz.
\figname~\ref{fig:compute}(f) present the modulus of the non-unitary matrix elements, which exhibit excellent agreement with theoretical values. Furthermore, the output intensities achieve a high effective bit resolution of $7.32$bit (coeff. = $0.99$, $\sigma^2=0.0125$).

\textbf{Neural network for image recognition}
To demonstrate the computational capabilities of \SiPhName for neural network applications, we implement a one-layer neural network on our processor and evaluate it using the Iris dataset, which comprises $150$ samples with $4$ numeric features across $3$ classes.
We conducted two experiments to validate our approach \figname~\ref{fig:compute}(o,p).
In the first experiment, we perform offline training of a unitary neural network structure that can be represented by the processor, achieving an inference accuracy of $94.67\%$.
After programming the trained phase shift values into the photonic processor, we obtain an online inference with an accuracy of $93.33\%$.
\figname~\ref{fig:compute}(g) presents the inference confusion matrix along with examples of the output intensity distributions for the Virginia, Verssicolor, and Setosa class(category $1$), respectively, demonstrating that the silicon-integrated programmable photonic processor can achieve a competitive inference performance to its electronic counterpart.
In the second one, we implement training and inference on the photonic platform.


\subsection{Signal processing for Optical I/O in AI clusters}
As the electrical I/O approaches its fundamental limitations in AI computing systems, optical I/O emerges as a promising alternative.
Silicon microring modulators (MRM), due to their wavelength selectivity, low power consumption, and compact footprint, have shown significant potential for high-speed signal transmission \cite{wade2020teraphy,dasforum}.
However, their performance is highly susceptible to thermal variations, which reduce the modulation depth and bit error rate (BER) of transmission links.
To achieve a high extinction ratio (ER) for modulated signals, while various electronic-based real-time feedback tuning techniques have been developed to maintain the wavelength alignment between the source and MRM resonance\cite{padmaraju2012thermal,li2014silicon,sun2018128}, it is, to our best of knowledge, the first exploration to utilize a programmable photonic processor in the control strategy, which process optical symbols with low latency and high power efficiency.

\begin{figure}
\centering
\includegraphics[width=\textwidth]{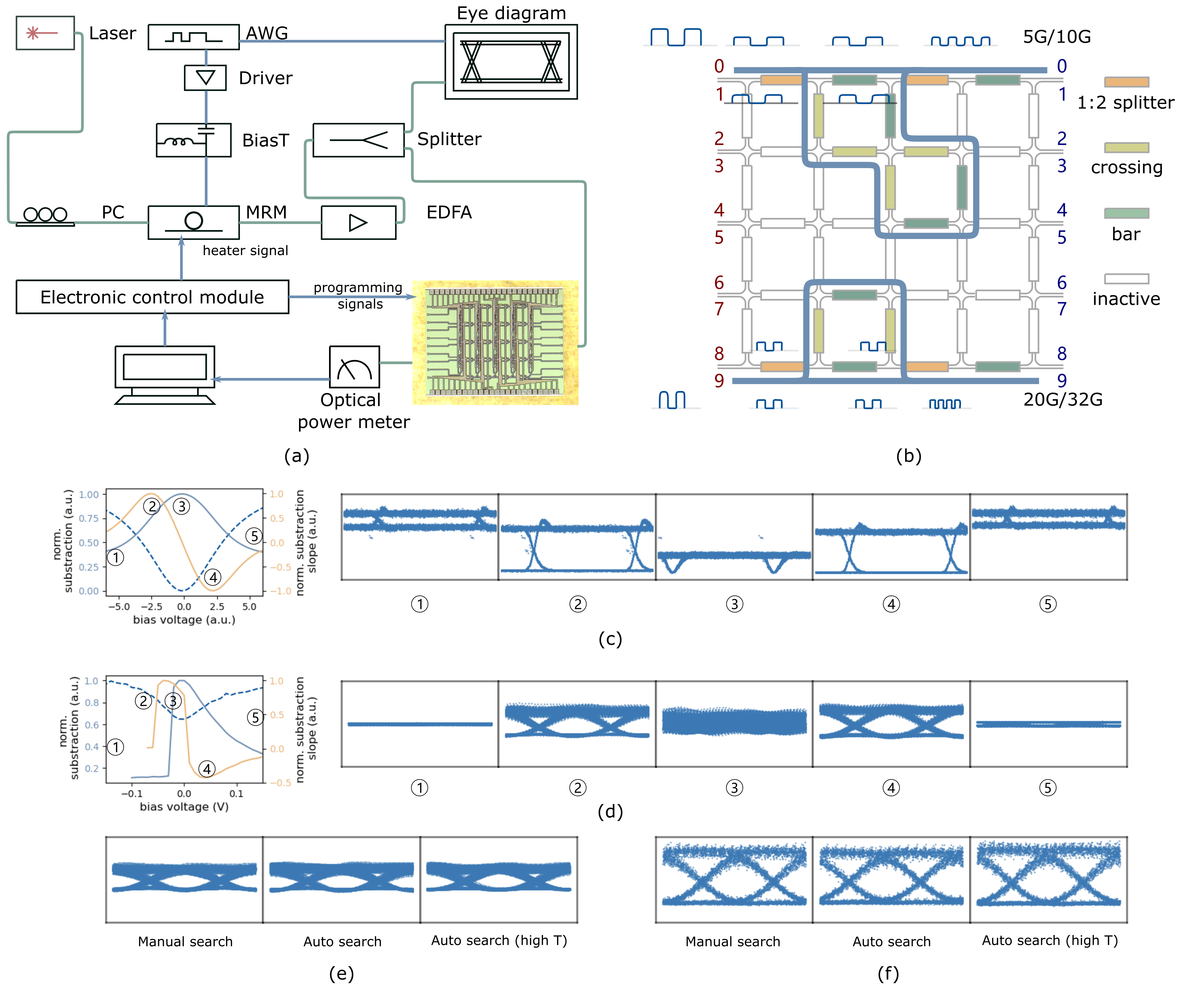}
\caption{Experiment system and results of \SiPhName to automatically lock wavelength for MRM in optical I/O.}
\label{fig:oio}
\end{figure}

A high ER corresponds to a significant amplitude difference between the pulse logic $`1'$ and $`0'$.
To obtain the amplitude difference, we configure the processor as a differentiator as in \figname~\ref{fig:oio}(a), which can complete the complex amplitude subtraction of signals at light speed.
The subtracted light is converted into an electronic monitor signal through photo-electronic conversion.
This signal is proportional to the square of the complex amplitude subtraction between adjacent symbols.
Note that at the microring resonance wavelength, the light phase of the modulated signals exists a $\pi-$ jump.
When the modulation signals are on the two sides of the resonance wavelength, it makes the complex amplitude subtraction the amplitude absolute value addition.
As the MRM resonance wavelength approaches and retreads from the laser source, the power of the electronic monitor signal will grow and decline.
When the MRM resonance wavelength is at the critical points of growth and decline, the modulated signal can obtain a high ER as in \figname~\ref{fig:oio}(c) that depicts the simulation results.
Therefore, we can utilize a micro-control circuit to read the electronic signal following the output electronic monitor signal and its variation and output the MRM heater-adjusting signal to increase or maintain a high ER.
The feedback loop is from the MRM output, \SiPhName-based differentiator, micro-control circuit, to the MRM heater.

We establish an experimental system as in \figname~\ref{fig:oio}(a).
By manually adjusting the heater voltage in the $10$G transmission system, the obtained eye diagrams, symbol subtraction and the subtraction slope, and MRM output power show good agreement with the simulation results as in \figname~\ref{fig:oio}(c).
This agreement verifies the principle correctness of the proposed method.
Furthermore, we implement the automatic approach for MRMs with $5$G and $32$G NRZ modulations under temperatures of $25^\circ C$ and $35^\circ C$, respectively.
Since the symbol widths vary with the baud rate, the appropriate delay differences in the photonic differentiator should be adjusted accordingly.
As shown in \figname~\ref{fig:oio} (b), the processor achieves this adaptation through programming.
The eye diagrams of the MRM working under the automatically locked wavelengths in different symbol rates and temperatures, which fit well with manually chosen ones, prove that the \SiPhName-based automatic locking hardware can align the MRM optimum working wavelength to the laser source.

\subsection{Optical switching in AI cluster}

\begin{figure}
\centering
\includegraphics[width=\textwidth]{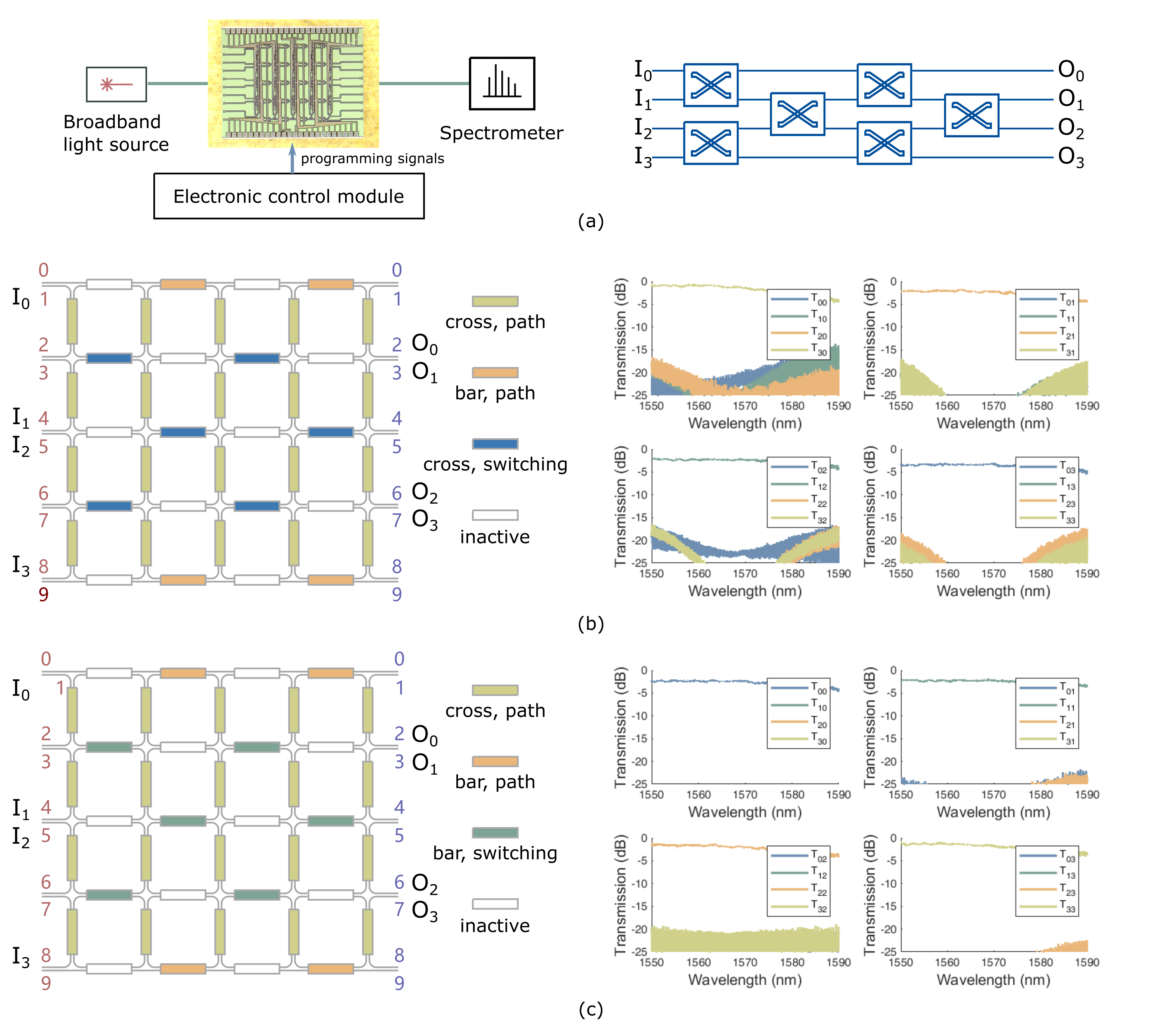}
\caption{Experiment system and results for optical switching implemented on \SiPhName.}
\label{fig:switch}
\end{figure}

To accommodate the rapidly growing data transmissions among multiple nodes in high-performance AI computing clusters, optical switching has been a promising technology because of its adaptive resource allocation, low latency, high bandwidth, and high power efficiency.
There are diverse MZI-based switching topologies, including the rearrangeable non-blocking N-stage planar and wide-sense non-blocking path-independent loss (PILOSS) structures\cite{8936941}.
Here, we realize a $4-$stage planar switching structure on the \SiPhName (\figname~\ref{fig:switch}(a)), which eliminates optical crossovers and provides rearrangeably non-blocking operation.
To validate the performance, we test the spectral characteristics for all measured optical paths with the experiment system as in \figname~\ref{fig:switch}(a).
The results demonstrate crosstalk leakage to other ports is at least $-20$dB and up to $-40$dB at the central wavelength $1560$nm, remaining under $-15$dB and $-20$dB across a bandwidth exceeding $20$nm for all-crossing and all-bar states, respectively (see \figname~\ref{fig:switch}(b,c)).
While the crosstalk merit is not satisfying, it could decrease by designing and manufacturing high-ER MZIs.
Additionally, the \SiPhName is limited in scale to realize the PILOSS structure, which can achieve a loss of uniformity across all paths. Future works will implement the PILOSS structure on the processor as in Supplementary Information Note 4 \figname~SI.2.


\subsection{Information security for AI clusters}
\begin{figure}
\centering
\includegraphics[width=\textwidth]{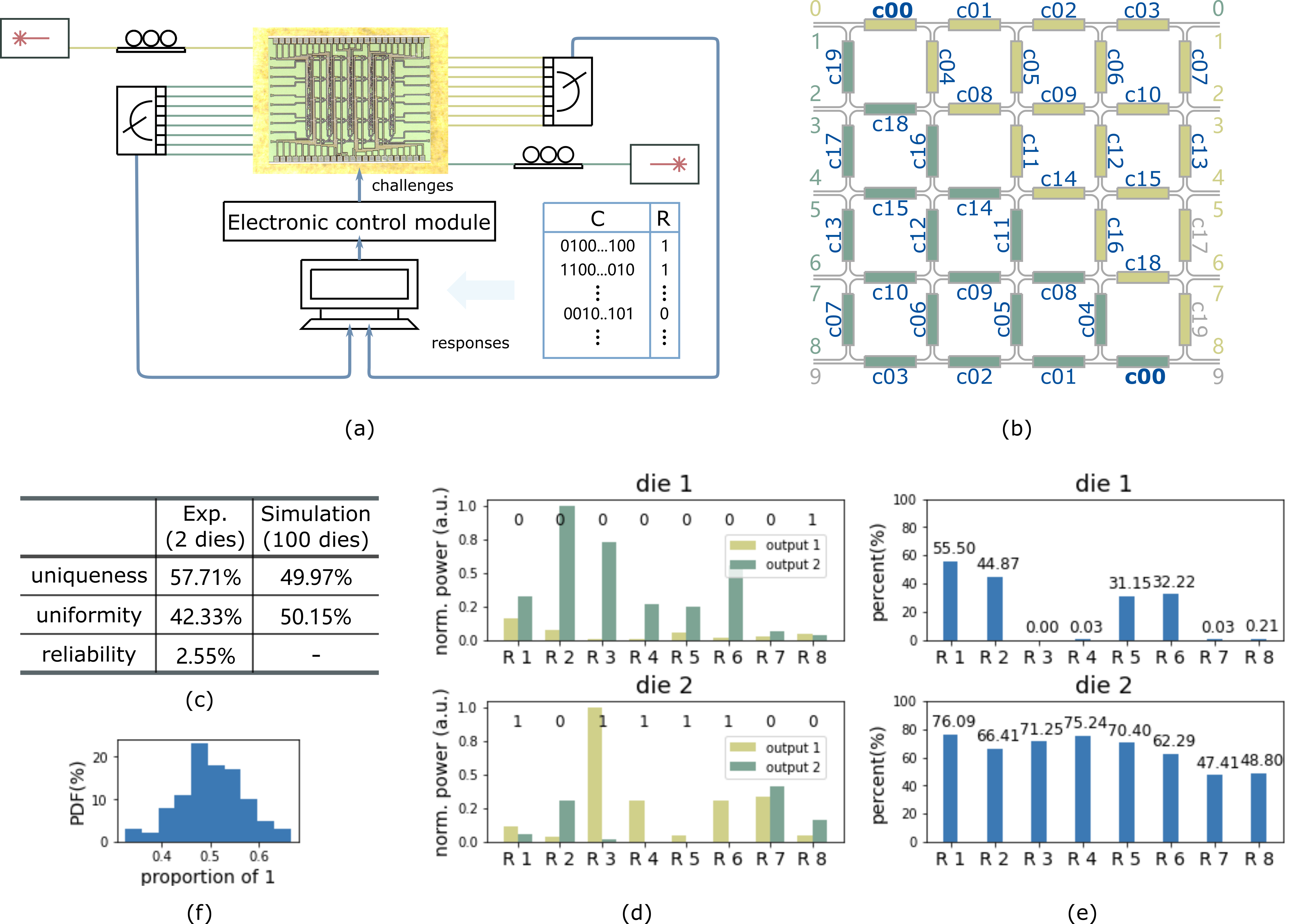}
\caption{Experiment system and results for optical PUFs implemented on \SiPhName.}
\label{fig:puf}
\end{figure}

The rapid development of AI computing and massive data transmission has increased the information security requirements \cite{wang2020recent}, where traditional security systems storing the secret keys in nonvolatile memory (NVM) are vulnerable to external attacks or require complex circuits.
PUFs, taking advantage of their inherent hardware randomness from the manufacturing process, can act as the 'fingerprint' in the computing systems \cite{gao2020physical}.
When a PUF is stimulated by an input challenge $C$, the output response $R = f(C)$, where $C$ and $R$ are multi-bit data and $f(\cdot)$ is determined by the PUF design structure and its manufacturing hardware.
Inspired by the arbiter PUFs with electronic integrated circuits, we propose a novel rotational-symmetric PUF structure design as in \figname~\ref{fig:puf} with two theoretical equal-power lights are injected into the diagonal-position input port $1$ and $2$.
Challenges are the programming voltages. 
$`1'$ being the high-level voltage and $`0'$ the low-level voltage, theoretically corresponding to cross and bar states, respectively.
Voltages are the same applied to the equal logical position of MZIs with the same indices but different colors as in \figname~\ref{fig:puf}.
Theoretically, since the structure is rotationally symmetric, light powers from the corresponding output ports are equal.
However, the output powers will deviate due to the process variations, the bit precision of the programming voltages, and environmental noises.
We define the response $r_i=1$ if the output port $o_{i,1}\ge o_{i,2}$, otherwise $r_i=0$, where $i$ is the $i-$th bit in the response $R$.

To assess a PUF, three merits are used, i.e., uniqueness (UQ), uniformity (UF), and reliability (RL) \cite{athanas2012embedded}.
The uniqueness is evaluated by the inter-die Hamming difference (HD), i.e.,
\begin{equation}
UQ = \frac{2}{N(N-1)}\sum_{i=1}^{N-1}\sum_{j=i+1}^N HD(R_i,R_j)/|R|,
\end{equation}
where $N$ is the total number of chip dies, $R_i$ and $R_j$ are responses to the die $i$ and $j$, and $|R|$ is the response bit length.
It represents the different responses from applying the same challenge to two PUF dies.
The theoretical UQ value is $50\%$.
Due to the chip number limits, we only test two dies with $128$ challenges by experiments.
The inter-die Hamming difference between the two dies is $57.71\%$.
Furthermore, we simulate $100$ dies, in each of which the initial length differences between two arms in every MZI follow a Gaussian distribution $\mathcal{N}(\mu=-0.08e^{-6},\sigma=0.11e^{-6})$ whose parameters are from the prototype processing system with the above described TPC framework.
The inter-die Hamming difference between $100$ simulated dies is $49.97\%$, demonstrating the PUF's uniqueness.

The uniformity is assessed by the proportion of $``0"$ and $``1"$  of the responses for one die with
\begin{equation}
UF = \frac{1}{M}\sum_{i=1}^M R_i/|R|,
\end{equation}
where $M$ is the number of different challenges applied and $R_i$ is the corresponding responses to the challenge $C_i$.
A higher uniformity, meaning randomness of challenge-response pairs, benefits applications such as key generation and authentication.
According to the definition, UF's optimal value is $50\%$.
The experimental results show that the averaged proportions of "$1$" for the two dies is $42.33\%$.
In addition, the average uniformity for $100$ simulated dies is $50.15\%$, demonstrating promising randomness.

Reliability is the reproduction of the same response from the same challenge under multiple measurements, evaluated by the intra-die Hamming difference
\begin{equation}
RL = \frac{1}{N}\sum_{i=1}^N\sum_{j=1}^{M} HD(R_i,R_{i,j})/|R|,
\end{equation}
where $N$ is number of challenge-response pairs,$M$ is the measurement times, and $R_i$ and $R_{i,j}$ are responses to the challenge $C_i$ under the reference and the measurement $j$.
We test the PUF $10$ times with $128$ challenges. The intra-die HD of their responses is $2.55\%$, which is close to $0\%$.
Noted,  the intra-die HD between the nominal and other temperature conditions increases to over $10\%$.
While this temperature sensitivity impacts the PUF reliability, it extends its functions to the PUF sensor and random number generations\cite{gao2020physical}. 
The experimental and simulation data demonstrate it is a promising structure for photonic PUF design.

\section{Discussion}\label{sec:discussion}

While the \SiPhName has achieved diverse functions, we still find the current limitations and propose potential improvements for future work. 

\textbf{Component Level Optimization}.
The current MZI, which employs one phase shifter, faces arm imbalance and high power consumption with the applied voltage ranging from $0$ to $2V_\pi$.
To address these concerns, another three phase shifters will be added to the MZI: one at the other arm parallel to the current one and two at the arms after the input ports.
The updated MZI configuration enables precise control of the two-arm phase difference, allows programming any phase shift within $V_\pi$ voltage, and realizes arbitrary complex unitary transformation for abundant matrix expression.
Furthermore, advanced design and manufacturing processes are required for the MZIs to achieve low loss and consistency in large-scale square mesh implementation.
Additionally, the current grating-based input/output couplers, which exhibit narrow bandwidth and high coupling loss, can be replaced with edge couplers to support high-speed modulation signals in optical computing and switching applications.

\textbf{Circuit Level Enhancement}.
While the waveguide lengths among the MZIs in the square mesh are designed equal, the waveguide lengths between the input/output gratings and the MZIs are not uniform and cannot be calibrated due to the lack of phase shifters.
This inconsistency leads to temporal misalignment among parallel input signals, particularly affecting computing precision and speed in optical computing applications.
Future designs will incorporate equal waveguide lengths and additional phase shifters for precise path control.

\textbf{System integration and automation.}
In the current \SiPhName, the electronic control module is a multi-port voltage source without control instruction storage and processing unit offered by the host computer running the TPC framework.
Therefore, the \SiPhName, consisting of multiple separated electronic and photonic devices, is bulky and cumbersome.
Besides, the topologies are pre-determined for different applications.
As the square mesh scales up, complex topologies are required.
For example, when four or eight differentiators are programmed in the programmable photonic processor simultaneously to control multiple MRMs, MZI allocation and path routing will become complicated.
Manual resource assignment is probably not the proper method. Furthermore, for fault tolerance, when some MZIs and paths are dead, other available MZI resources and input/output ports should be rapidly and automatically toggled to ensure system operations.
Therefore, an automated compilation flow is significant for the large-scale processor in our next-stage work.
In future work, we can integrate the compilation flow and multi-port voltage source in one System-on-chip (SoC)-enabled FPGA.
More importantly, it can be packaged with the processor to enable tighter integration.

\textbf{Future Vision: Photonic AI Computing Cluster}.
The current AI computing cluster is electronic-dominant, primarily centered on electronic digital chips such as GPUs, CPUs, and network switchers.
While the programmable photonic processor has the potential to provide high-speed and high-power efficiency signal processing, only implementing them in small parts of the current AI computing clusters requires numerous optoelectronic and analog/digital conversions to adapt to the large number of existing parts that process electronic signals.
It will introduce a large amount of power overhead and speed bottlenecks.
Therefore, we expect a photonic-based or hybrid optoelectronic AI computing cluster that requires minimal signal conversions.
In this novel architecture, the photonic SoC will complete numerous functions.
Before designing and fabricating the ASPIC, a programmable photonic processor can be programmed to different functions for rapid application verification, interconnected through an efficient data interface, forming a complete system optimized for high-speed and energy-efficient computing and transmission.
We believe this architectural innovation is on the horizon and will significantly advance the field of AI computing.
The development of the photonic processing system is indispensable.

\section{Methods}\label{sec:method}

\subsection{Design, fabrication and packaging of the silicon-integrated programmable photonic processor}\label{ssec:photonic}
The photonic processor occupies a square footprint of $3.8\times3$ mm$^2$.
It mainly consists of three kinds of components: the grating couplers, the MZIs, and the electronic pads.
$20$ grating couplers equally distributed at the two opposite edges spacing $222.22\mu$m.
The $40$ MZIs connect as a square mesh with $450$nm-width silicon waveguides.
Each MZI employs a  $100\mu m$-length heater as the phase shifter controlled by a pair of electronic pads.
The heaters in one column share one ground pad.
Therefore, on the photonic processor exist $49$ pads, among which $24$  and $25$ are distributed on the other two opposite edges with the spacing of $152\mu m$ and $145\mu m$, respectively.
The photonic processor is fabricated on a silicon-on-insulator (SOI) wafer, which has a $2.2\mu m$-thick oxide (SiO$_2$) cladding layer, a $220 nm$-thick silicon(Si) layer, and a $2\mu m$-thick buried thermal oxide (SiO$_2$) layer.
The grating couplers are interfaced through two fiber arrays.
The electronic pads are connected to a PCB with wire bonding.

\subsection{Prototype system construction}\label{ssec:prototype}
In this section, we will describe the experimental system constructions and the utilized devices for the above-mentioned applications.

\textbf{Computing acceleration in AI cluster}.
We establish two experiment systems to verify the computing acceleration functions implemented in the \SiPhName.
One is a low-speed experiment system to validate the unitary and non-unitary matrix expressions for the processor.
The other is a high-speed experiment system to demonstrate the computing performance of the \SiPhName.
In the former experiment, a multi-channel laser source (SOUTHERN PHOTONICS, TLS150-20) provides the $8$ independent lights for the bi-direction unitary matrix-vector multiplications ($4$ for each direction) and $6$ for the non-unitary ones.
The light wavelengths are set to $1560$ nm, respectively, and their powers are set to $16$mW or $0$mW for the matrix expression test and other corresponding values to express the Iris data for the NN test.
The multiplied optical signals are detected by a multi-channel power meter (KEYSIGHT, N7745A).
In the high-speed experiment, the carries are modulated via the intensity modulators (NOEIC, MZ1550-LN-40) by the AWG (KEYSIGHT, M8194A), and the multiplied signals are detected by the PD and sampled by the oscilloscope (KEYSIGHT, UXR0594A).

\textbf{Signal processing for optical I/O}.
To verify the \SiPhName functions in signal processing to realize wavelength-locking for MRMs, an experimental system, as shown in \figname~\ref{fig:oio}, is established.
A single-wavelength laser (SOUTHERN PHOTONICS, TLS150-20) is set to $1555$nm and $16$mW as the carrier.
The light is injected into an MRM and modulated by the RF signal generated from an AWG (KEYSIGHT, M8194A).
The modulated signal is then amplified by an EDFA (Amonics, AEDFA-CL-20-R-FC) split into two paths, one to eye diagram (KEYSIGHT, DCA-M N1092A) to demonstrate the signal quality and the other to the \SiPhName-based signal processing system for optimum bias point searching, in which the device for diffraction signal detection is the power meter (KEYSIGHT, N7745A).
EDFAs, compensating for the coupling loss, can be removed when the photonic components for the automatic locking are integrated with MRMs in one chip.

\textbf{Optical switching in AI cluster}.
To demonstrate the \SiPhName performance as an optical switcher, we use a broadband light source (Realphoton, ASE-B-F-CL-50-S-FA) and an optical spectrometer (YOKOGAWA, AQ6370D) to observe the transmission spectrums between any two input and output ports.

\textbf{Information security for AI cluster}.
In this prototype system, two single-wavelength lasers (SOUTHERN PHOTONICS, TLS150-20) are set to $1560$nm and $1560$nm, respectively, to avoid power fluctuation from the optical heterodyning between two lights.
The two laser powers are around $12.5$dBm and are carefully calibrated by the polarization controllers (PC) between the laser source and the photonic processor to maintain the injection power equality according to the PUF design principle.
The host computer generates the challenge bit, stimulates the PUF via the multi-channel voltage source (NOEIC, MCVS-128C), and reads and compares the PUF output powers through the multi-channel power meter (KEYSIGHT, N7745A) to obtain the responses.

\end{document}